\documentclass[twocolumn]{aastex62}
\graphicspath{{./}{figures/}}

\received{}
\revised{}
\accepted{}
\submitjournal{ApJ}

\shorttitle{A new tidal structure in the LMC}
\shortauthors{}

\begin{document}

\title{Discovery of NES, an extended tidal structure in the North-East of the Large Magellanic Cloud}

\correspondingauthor{Massimiliano Gatto}
\email{massimiliano.gatto@inaf.it}

\author[0000-0003-4636-6457]{Massimiliano Gatto}
\affiliation{INAF-Osservatorio Astronomico di Capodimonte, Via Moiariello
 16, 80131, Naples, Italy \\}
\affiliation{Dept. of Physics, University of Naples Federico II, C.U. Monte Sant'Angelo, Via Cinthia, 80126, Naples, Italy\\}

\author{Vincenzo Ripepi}
\affiliation{INAF-Osservatorio Astronomico di Capodimonte, Via Moiariello
 16, 80131, Naples, Italy \\}

\author{Michele Bellazzini}
\affiliation{INAF-Osservatorio di Astrofisica e Scienza dello Spazio, Via Gobetti 93/3, I-40129 Bologna, Italy \\}

\author[0000-0001-7958-6531]{Crescenzo Tortora}
\affiliation{INAF-Osservatorio Astronomico di Capodimonte, Via Moiariello
 16, 80131, Naples, Italy \\}

\author{Monica Tosi}
\affiliation{INAF-Osservatorio di Astrofisica e Scienza dello Spazio, Via Gobetti 93/3, I-40129 Bologna, Italy \\}

\author{Michele Cignoni}
\affiliation{INAF-Osservatorio Astronomico di Capodimonte, Via Moiariello
 16, 80131, Naples, Italy \\}
\affiliation{Physics Departement, University of Pisa, Largo Bruno Pontecorvo, 3, I-56127 Pisa, Italy\\}
\affiliation{INFN, Largo B. Pontecorvo 3, 56127, Pisa, Italy\\}

\author{Giuseppe Longo}
\affiliation{Dept. of Physics, University of Naples Federico II, C.U. Monte Sant'Angelo, Via Cinthia, 80126, Naples, Italy\\}



\begin{abstract}

We report on the discovery of a new diffuse stellar sub-structure protruding for $>5$\degr~from the North-Eastern rim of the LMC disc. The structure, that we dub North-East Structure (NES), was identified by applying a Gaussian Mixture Model to a sample of strictly selected candidate members of the Magellanic System, extracted from the Gaia EDR3 catalogue.
The NES fills the gap between the outer LMC disk and other known structures in the same region of the LMC, namely the Northern tidal arm (NTA) and the Eastern sub-structures (ES).
Particularly noteworthy is that the NES is placed in a region where N-body simulations foresee a bending of the LMC disc due to tidal stresses induced by the MW. 
The velocity field in the plane of the sky indicates that the complex of tidal structures in the North-Eastern part of the LMC, including NES, is subject to coherent radial motions. Additional data, as well as extensive dynamical modeling, is required to shed light into the origin of NES as well as on the relationships with the surrounding substructures.
\end{abstract}

\keywords{}


\section{Introduction} \label{sec:intro}

The Large Magellanic Cloud (LMC) is the largest dwarf satellite of the Milky Way (MW) and along with its smaller companion, the Small Magellanic Cloud (SMC), represents a very close example of a physical system in an ongoing tidal stripping process. 
Some signatures of their mutual interaction are very prominent in the sky, such as the Magellanic stream, a cloud of HI gas that extends for more than 180\degr~in the Southern hemisphere \citep[][]{Putman2003,Bruns2005,D'onghia&Fox2016}, or its counterpart, the Leading arm \citep[][]{Putman1998}. The Magellanic Clouds (MCs) are also connected by the Magellanic bridge, a stream likely originated as a consequence of the last close encounter between the MCs \citep[][]{Zivick2018}, made of HI gas \citep[][]{Hindman1963}, young \citep[e.g.,][]{Harris-2007,Skowron-2014,Noel-2015,Mackey-2017} and intermediate-old stars \citep[][]{Bagheri2013,Carrera2017,Jacyszyn-Dobrzeniecka2017}.
Because of their proximity, namely $\sim 50$~kpc the LMC \citep{degrijs-wicker-bono-2014} and $\sim 60$~kpc the SMC \citep[][]{degrijs&bono-2015}, they offer a unique opportunity to investigate in great detail the physical effects of a three-body encounter, among the LMC, SMC, and the MW.
According to N-body simulations, the expected outcome of these interactions is a wealth of complex and extended tidal features, that may be very difficult to detect because of their very low surface brightness \citep[e.g.,][]{Bullock2005,Cooper-2010}.
Moreover, according to simulations, the origin and the present-day shape of these features require not just tidal effects, but also hydro-dynamical effects of interaction between the MCs \citep[e.g.][and references therein]{Wang2019}.\par
In the last decade, several of these structures have been discovered, thanks to modern deep panoramic surveys
\citep[e.g,][]{Mackey-2016,Belokurov&Koposov2016,Pieres2017,Choi2018a,mackey2018,Massana2020,ElYoussoufi-2021}.
These discoveries aided to decipher the interaction history of the MCs, whose morphology and kinematics have been shaped in the last Gyrs by their mutual attraction and by the gravitational interference induced by the MW \citep[e.g.,][]{Besla-2012,Belokurov&Erkal2019}.
Even more valuable, the Early Data Release 3 (EDR3) of the {\it Gaia} mission \citep[][]{Gaia2016,GaiaBrown2021} made it possible to kinematically characterize the already known tidal substructures \citep[e.g.,][]{Zivick2019,Schmidt-2020,Omkumar-2021,Cullinane-2022,James-2021,Piatti-2022}, to unveil the response of the MW halo to the LMC \citep[e.g.,][]{Belokurov-2019,Garavito-Camargo-2019,Conroy-2021,Vasiliev-2021}, and also to discover new stellar streams \citep[][]{Belokurov&Erkal2019,DeLeo2020,Grady-2021,Petersen-2021}.\par
Amongst the numerous sub-structures revealed in the last years, the arm-like stellar over-density to the North of the LMC, currently known as Northern Tidal Arm (NTA), discovered by \citet{Mackey-2016} thanks to the analysis of the public Dark Energy Survey (DES), is particularly prominent.  
This sub-structure was traced in all its extension by \citet{Belokurov&Erkal2019} using red giant branch (RGB) stars selected by means of {\it Gaia} Data Release 2 \citep[DR2,][]{Gaia2018}.
Different authors ran N-body simulations of a MW-LMC interaction to explain the origin of this feature \citep[][]{Mackey-2016,Belokurov&Erkal2019,Cullinane-2022}.
For example, \citet{Belokurov&Erkal2019} simulated the evolution of the LMC particles in the last 1 Gyr in the presence of the only MW or SMC, and also under the influence of both galaxies. They also sampled the simulation outcomes with different LMC and SMC mass values.
They concluded that the MW is the main actor producing the bending of the Northern part of the LMC disc, but a combination of both the MW and the SMC is needed to originate the ensemble of tidal substructures observed around the LMC.
\citet{Cullinane-2022} also produced a suite of dynamical simulations of MW-LMC or MW-LMC-SMC interactions, letting them evolve during the last 1 Gyr, and treating each system as a particle sourcing potential in order to consider the response of the MW motion to the LMC.
Similarly to \citet{Belokurov&Erkal2019}, they inferred that the tidal force of the MW is the primary responsible of the LMC Northern bending.
It is worth noticing that all these simulations predict a more diffuse twisted stream, located towards the North-East of the LMC, rather than a thin and horizontal arm as observed so-far.\par
In this paper, we use {\it Gaia} EDR3 results to select stars belonging to the intermediate and old population of the LMC to trace the very low-surface brightness regions in the outskirts of the LMC. This procedure allows us to discover a more diffuse stellar structure protruding from the outer disk of the LMC in the North-East direction, right below the NTA and in the same position predicted by existing dynamical models \citep[][]{Mackey-2016,Belokurov&Erkal2019,Cullinane-2022}.\par


\section{Sample selection}
\label{sec:sample}

\begin{figure}
    \includegraphics[width = \hsize]{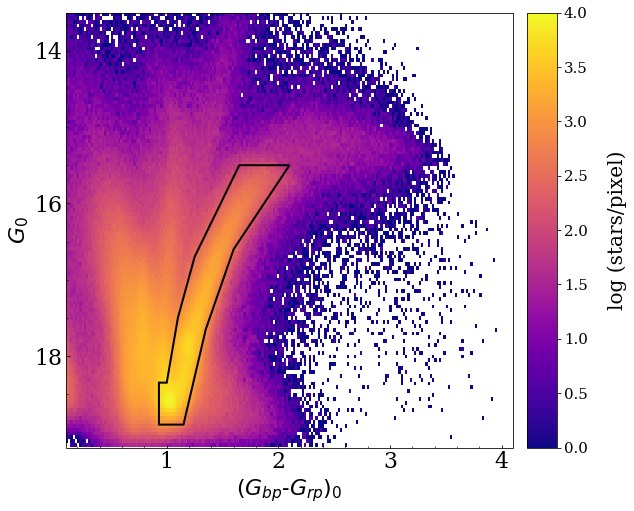}
    \caption{CMD of all stars retrieved with the Gaia query, corrected for reddening. The black polygon defines the RGB+RC star selection.}
    \label{fig:cmd_sel}
\end{figure}

To carry on a preliminary selection of stars potentially belonging to the LMC, we performed a query to the Gaia database\footnote{https://gea.esac.esa.int/archive/} to retrieve astrometric and photometric information from {\it Gaia} EDR3 \citep[][]{GaiaBrown2021}. In particular, we confined our research in a sky region defined by the Galactic longitude $200 \leq l \leq 360$ and Galactic latitude $-90 \leq b \leq -10$, in order to avoid likely MW disc stars which can cause significant contamination to the sample. Moreover, we restricted the query to objects whose PMs are within 3$\sigma$ from the mean LMC and SMC PMs as measured by \citet{Luri-2021}. Following a procedure similar to that described by \citet{Luri-2021}, we minimized contamination from MW foreground stars, by requiring that $\varpi \leq 3~\sigma_{\varpi}$, where $\varpi$ and $\sigma_{\varpi}$ are the parallax and the parallax uncertainties, respectively. The latter selection allowed us to exclude sources not compatible with the LMC distance.
In addition, we filtered out the stars having errors larger than 0.2 mas yr$^{-1}$ in each of the two PM components and the sources with ${\rm RUWE}\footnote{Re-normalised unit weight error, see Sect.~14.1.2 of ``Gaia Data Release 2 Documentation release 1.2''; https://gea.esac.esa.int/archive/documentation/GDR2/} > 1.3$ ${\rm astrometric\_excess\_noise\_sig} > 2 $ to limit our sample to objects with good PMs accuracy and reliable astrometric solutions (see App.~\ref{app:query} for a reproduction of the query). 
This query yielded a total of $4,363,722$ stars.\par
We used the photometry provided by {\it Gaia} to select both RGB and  red clump (RC) stars in the colour-magnitude diagram (CMD), using $G$, $G_{BP}$, $G_{RP}$ {\it Gaia} filters.
We focused only on the stars belonging to the intermediate-old populations, because they are known to be distributed at large distances from the LMC centre \citep[e.g.,][]{Luri-2021,ElYoussoufi-2021}. 
As a first step, we corrected the photometry for the reddening, to avoid that highly extincted MW disk stars could contaminate our sample. To this end, we adopted the reddening maps by \citet{Skowron-2021} that provided extinction values in the inner regions of the LMC and SMC, i.e. $\sim 6\degr$ and $\sim 5\degr$ from the LMC and SMC centre, respectively. Outside these regions, we sampled the reddening maps by \citet{SFD98} by means of the python package {\tt DUSTMAPS} \citep[][]{Dustmaps}. We calculated the extinction in the {\it Gaia} filters through the relations provided by \citet{Wang&Chen2019}. Using the dereddened photometry, we finally selected bona-fide RGB and RC stars by including in our sample all the stars inscribed within the polygon 
shown in Fig.~\ref{fig:cmd_sel} (see also App.~\ref{app:query}). 
This procedure yielded a total of 
$1,759,796$ stars, that after a quick visual investigation about their sky positions we deduced are a mix of LMC and SMC stars, with a still not negligible contamination by MW stars (see next section).
To further remove the MW noise, it is useful to adopt a new reference frame centred on the LMC, in which the new PMs represent the motion of a star in the plane of the LMC disc.
To this aim, we took advantage of the transformation laws reported in \citet{Luri-2021} (see their Eqs. 5 and 6) to define the new coordinates $\xi, \eta$, the new PMs $\dot \xi, \dot \eta$, which represent the coordinates and PMs of a star moving on the flat disc defined by the plane of the LMC\footnote{We adopted the inclination and position angle values reported in \citet{Luri-2021} to correct for the orientation effects.}.
In this new reference system we also calculated the radial and tangential velocities $v_R, v_{\phi}$, and the third component of the angular momentum $L_z$.
The quantity $L_z$ is particularly useful to further clean our sample. In particular, we estimated the average LMC $L_{z,LMC}$ by using all the stars within $8\degr$ from the LMC centre and filtered out all the stars that did not satisfy the condition $L_{z,LMC} - 5\sigma_{L_{z,LMC}} < L_z < L_{z,LMC} + 5\sigma_{L_{z,LMC}}$, where $\sigma_{L_{z,LMC}}$ is the standard deviation of ${L_{z,LMC}}$. This cut yielded a total of $1,452,155$ stars.

\section{Selection of LMC stars through machine learning techniques}
\label{sec:gmm}

\begin{figure}
    \includegraphics[width = \hsize]{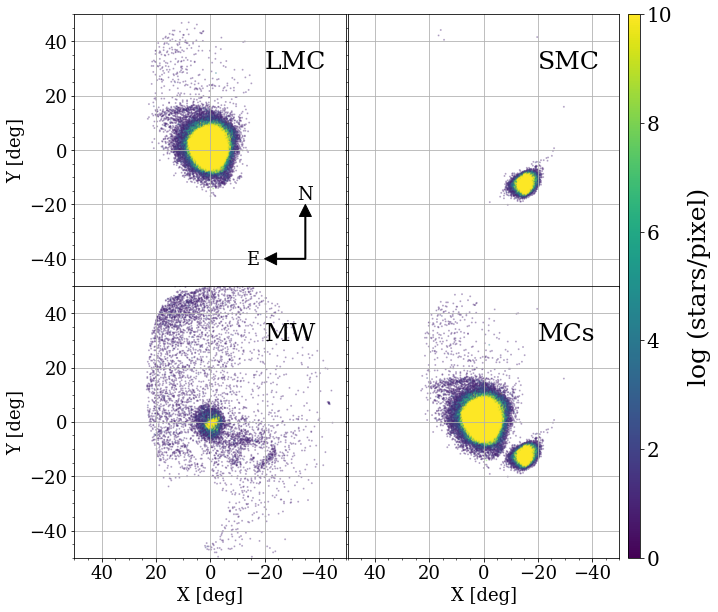}
    \caption{Spatial density of the stars belonging to the LMC, SMC, MW and MCs together, respectively, in an orthographic projection centred on the LMC. Each pixel has a resolution of $0.2\degr \times 0.2\degr$.}
    \label{fig:gmm_output}
\end{figure}

To discern stars belonging to different stellar sub-components we used a Gaussian Mixture Modelling (GMM)\footnote{We used the library available within scikit-learn python package \citep[][]{scikit-learn}}, which estimates the statistics of the Gaussians of each underlying single stellar sub-population and, in turn, evaluates for each star the probability of belonging to these sub-populations.
The GMM is an unsupervised clustering technique which needs only one parameter, i.e the number $N$ of finite Gaussian functions that describes the $N$ sub-populations within an overall population. We run the GMM with $N=3$ Gaussian to model the three stellar components of our sample: the LMC, the SMC and the MW. 
As input for the GMM we adopted six parameters\footnote{This set of parameters ensures a better separation of the three stellar components. The stellar structures we discuss in this work are clearly visible also with any other choice of input parameters anyhow.}: the classical PMs ($\mu_{\alpha}$, $\mu_{\delta}$); the parallax $\varpi$, the radial and tangential velocities defined in the new reference frame $v_R, v_{\phi}$; and the de-reddened colour of the stars $(G_{BP}-G_{RP})_0$.
Finally, we assigned to each stellar sub-component, i.e. LMC, SMC and MW, only stars having at least $99\%$ of probability to belong to one of the three galaxies, while the remaining objects are filtered out. After this step our total sample is reduced to $1,290,166$ stars.\par

\section{Results}
\label{sec:results}

\begin{figure}
    \centering
    \includegraphics[width=\hsize]{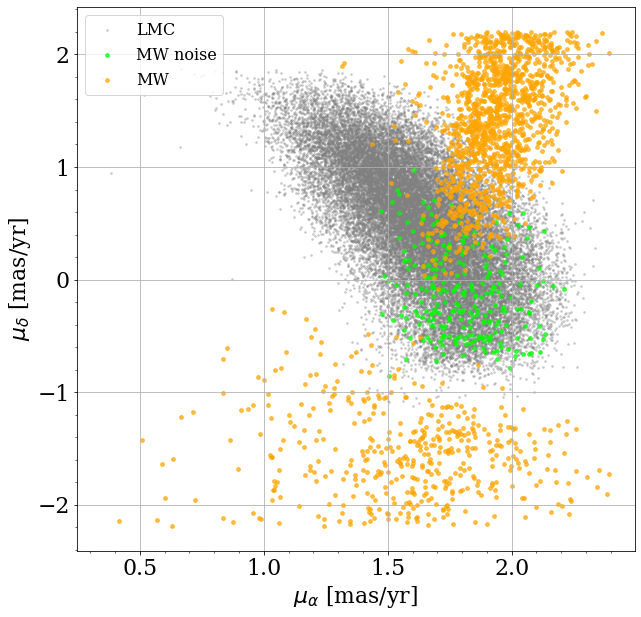}
    \caption{PM space of LMC disc stars beyond $8\degr$ (gray points) along with the MW noise (green points) and MW stars in the same region of the MW noise (orange points).}
    \label{fig:pm_space}
\end{figure}

Table~\ref{tab:gmm_stat} lists the statistics of the three Gaussians derived for each sub-population by the GMM, as well as the number of stars assigned to each component. 
It can be seen that the MW star residuals were $\sim1.3\%$, while the percentage of SMC stars raises to more than $5\%$, as expected because of the large overlapping of their stars on the CMD. 
The average PMs of the MCs are in very good agreement with those estimated by \citet{Luri-2021} using a selection of RGB stars, e.g. $\mu_{\alpha} = 1.77 \pm 0.44~{\rm mas~yr^{-1}}$; $\mu_{\delta} = 0.33 \pm 0.63~{\rm mas~yr^{-1}}$ for the LMC and $\mu_{\alpha} = 0.71 \pm 0.36~{\rm mas~yr^{-1}}$; $\mu_{\delta} = -1.22 \pm 0.29~{\rm mas~yr^{-1}}$ for the SMC, making us confident about the reliability of the GMM technique.\par
\begin{table*}[]
    \centering
    \begin{tabular}{lccccccc}
        Pop. & N & $\mu_{\alpha}$ & $\mu_{\delta}$ & $v_r$ & $v_t$ & $\omega$ & $(G_{BP}-G_{RP})_0$\\
        \hline
        LMC & 1202727 & $1.79 \pm 0.26$ & $0.39 \pm 0.45$ & $0.26 \pm 0.22$ & $-0.01 \pm 0.22$ & $-0.003 \pm 0.110$ & $1.2 \pm 0.2$\\
        SMC & 70856 & $0.80 \pm 0.24$ & $-1.10 \pm 0.10$ & $0.09 \pm 0.09$ & $0.50 \pm 0.50$ & $-0.024 \pm 0.113$ & $1.2 \pm 0.2$\\
        MW & 16583 & $1.85 \pm 0.62$ & $0.30 \pm 0.80$ & $0.25 \pm 0.56$ & $-0.01 \pm 3.64$ & $0.040 \pm 0.199$ & $1.1 \pm 0.1$\\
        \hline
        
    \end{tabular}
    \caption{Mean and standard deviations of the Gaussians estimated through the GMM.}
    \label{tab:gmm_stat}
\end{table*}
Figure~\ref{fig:gmm_output} shows the allocation of each star to one of the three components, displayed through an orthographic projection centred on the LMC \citep[we used Eq. 1 of][]{Luri-2021}. An inspection of the figure reveals that the GMM technique was able to properly separate the LMC from the MW and the SMC, with two exceptions. 
The first is the wrong assignment of LMC stars to the MW. 
These stars are mainly located within the LMC bar and we speculate that the more chaotic kinematics \citep[see e.g.][]{Luri-2021} of the central bar misleads the GMM so that it incorrectly associates these stars to the MW. Furthermore, the $v_{\phi}$ values within $4\degr-5\degr$ to the LMC centre is different with respect to those measured in the outer regions \citep[see e.g.,][their Fig.~14]{Luri-2021}.
The second is a cloud of stars ($\simeq 0.02\%$ of the stars assigned to the LMC), that we call ``MW noise'' hereafter, located between $20-50\degr$ North of the LMC (top-left panel) which we speculate are MW misclassified stars.
To confirm this hypothesis, we plotted in Fig.~\ref{fig:pm_space} the PMs of LMC stars placed beyond $8\degr$ from the LMC centre (gray points), of the MW noise (green points) and the stars assigned by the GMM to the MW situated in the same position of the MW noise (yellow points). The position and PM distribution of the latter population suggest that the MW noise is likely a tail of the MW sample that is being misclassified because of a significant overlap with the corresponding distributions of genuine LMC stars.\par
Regarding the stars assigned to the SMC (top right sub-panel in Fig.~\ref{fig:gmm_output}), besides its main body it is also visible the Small Magellanic Cloud Northern Over-Density (SMCNOD) discovered by \citet{Pieres2017}, which attests the ability of the GMM at revealing also very low-surface brightness structures \citep[][estimated $\mu_V$ = 31.2 mag arcsec$^{-2}$ for the SMCNOD]{Pieres2017}

\section{Discussion}
\label{sec:discussion}

\begin{figure}
    \centering
    \includegraphics[width=\hsize]{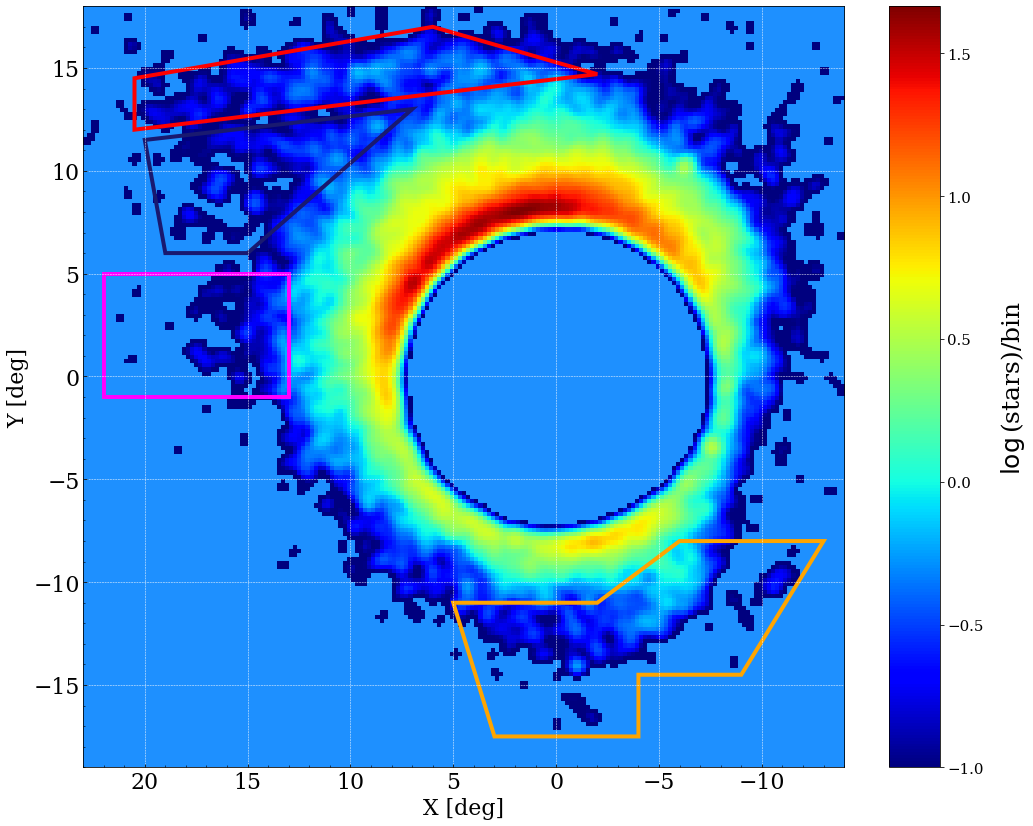}\\
    \caption{Zoom-in of the stars assigned to the LMC by the GMM in a density plot with a bin size of 0.2\degr $\times$ 0.2\degr, smoothed with a Gaussian function with $\sigma = 0.3$\degr. We masked the central 8\degr regions to enhance the distribution of stars in the outer regions.}
    \label{fig:gmm}
\end{figure}

In this section we discuss separately the morphology and velocity fields of the outer LMC substructures unveiled by the GMM algorithm.

\subsection{Morphology}

The spatial distribution of the stars selected as LMC members by the GMM is shown in Fig.~\ref{fig:gmm}. We only displayed objects located beyond 8\degr ~from the LMC centre to better visualize the outer regions (similar density plots with different choices of the bin size and/or smoothing width are displayed in Fig.~\ref{fig:kde_LMC}).
The morphology of the LMC towards the Galactic disk (i.e., North-East direction) is strongly disturbed as witnessed by several sub-structures emanating from the disc edge.
The most noticeable one is the already introduced NTA which is identified by a red polygon in the figure. 
It is also possible to discern some elongated features towards the East, up to $(X, Y) \simeq (20\degr, 0)$, that we dubbed Eastern Substructure (ES hereafter, highlighted with a magenta rectangle in Fig.~\ref{fig:gmm}, which represent the Eastern substructure I \citep[][]{deVaucouleurs-1955} and the recently unveiled Eastern substructure II \citep[][]{ElYoussoufi-2021}.
In addition, the GMM-selected sample of very likely LMC stars also reveals the ensemble of Southern substructures already discovered by \citet{Belokurov&Erkal2019} and \citet{mackey2018} that we collectively call S-Sub and highlight with an orange polygon.\par
The figure also reveals a diffuse sub-structure protruding from the outer LMC disc and extending at more than 20 deg from the LMC centre and placed in between the NTA and the ES. This feature, indicated with a blue polygon in Fig.~\ref{fig:gmm}, represents, as far as we know, a still undiscovered tidal feature, which we call North-Eastern Structure (NES).
The NES presents a horizontal over-density of stars (``Finger'' hereafter) at $Y \simeq 11\degr-12\degr$ which mimics the NTA at about $5\degr$ South of this feature and a more diffuse component which departs from the outer disk at $Y \simeq 5\degr-7\degr$ deg and connects itself with the Finger at about $(X, Y) \simeq (15\degr-20\degr, 10\degr)$.\par
\begin{figure}
    \centering
    \includegraphics[width=\hsize]{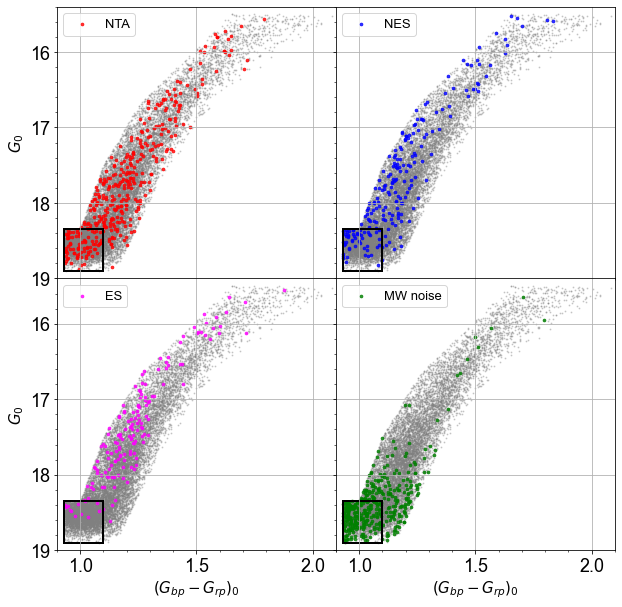}\\
    \caption{CMD of stellar sub-structures described in the text (labeled in the top left corner of each sub-panel), superimposed with an LMC outer disc stellar population located in the North-East side. The black square defines the RC.}
    \label{fig:cmd_pop}
\end{figure}

Figure~\ref{fig:cmd_pop} shows the CMD of the stars belonging to the NTA, the NES, the ES and the MW noise (highlighted with different colours), superposed with the CMD of LMC stars at its North-East side and beyond $9\degr$. The NTA (top-left panel) and the NES (top-right panel) have an RC-RGB similar to the underlying population of the outer LMC disc. This occurrence strengthens the hypothesis that the NES is composed by stars tidally stripped from the outer LMC disc, as the NTA. To be more quantitative, the ratio ($f$) between the number of RC (indicated with the black polygon in the figure) and RGB stars for the LMC outer disc, the NES and the NTA are very similar, being $f = 0.34 \pm 0.01$, 
$f = 0.29 \pm 0.05$ and $f = 0.28 \pm 0.05$, respectively.
The ES, instead, shows an almost total lack of RC stars. Such lower fraction for the ES might be due to issues with the completeness and/or reddening estimates, however investigating this occurrence is beyond the purpose of this work. Finally, the MW noise (bottom right panel) contains stars mostly having $G_0 > 18$, reinforcing the hypothesis that they are MW misclassified stars.\par
To further strengthen the reality of the NES, we point out that some hints of its presence were already visible in other works that exploited the {\it Gaia} catalogue. For example, the Finger is barely visible in the density maps constructed with {\it Gaia} EDR3 in \citet{Luri-2021} (their fig. 17), but the NES is more visible in fig. 2 of  \citet{Grady-2021}, which shows a selection of 226,119 RGB stars in both the LMC and SMC from the {\it Gaia} DR2.
The main scope of their work was to obtain photometric metallicity estimations and an overall picture of the PMs of their sample of stars rather than discover new undetected substructures in the faint outer regions of the LMC, therefore they also did not need to remove the minor MW component from their sample. The GMM adopted in this work cleaned the RGB stars from the MW contamination, yielding an almost pure catalogue of LMC stars, hence particularly enhancing the NES visibility.\par
To investigate the origin of the NES we inspected the recent results obtained using N-body simulations of the interaction between the LMC, SMC and MW. 
\citet{Mackey-2016} ran N-body MW-LMC simulations to understand the physical origin of the NTA. They demonstrated that the tidal interaction with the MW can warp and distort the outer LMC disk, generating two spiral-like patterns, one twisted towards the North-East and the other towards the South-West direction. Their results strongly indicate that the NTA originated by the tidal interaction of the LMC with the MW. Similar structures have been predicted by \citet{Belokurov&Erkal2019} N-body models. In particular, simulations of the LMC-SMC-MW or of the MW-LMC interactions, under some initial conditions\footnote{Of course, N-body simulations are strongly dependent on the initial conditions, in particular the galaxy masses involved, which are not known accurately yet.} foresee both a leading and trailing spirals, stretched in the Southwest-Northeast direction (see their Fig. 4).  
They speculated that the NTA originated from the gravitational interaction with the MW, but a combined influence of the MW and SMC is necessary to explain all the features observed in the South of the LMC.
Interestingly, these models also forecast, as a distinct and separated feature from the NTA, the bending of the entire Northern outer disc of the LMC caused by the MW gravitational pulling. Therefore, besides a thin horizontal stellar over-density, one should observe a more diffuse and wide stellar structure which should represent the distorted Northern LMC disk.
However, observational works until now did not disclose any diffuse stellar over-density in the North, but only the thin NTA. 
Morphologically, the NES is in the position expected from the quoted dynamical models and we can hypothesize that it might have had the same origin of the NTA. In this scenario it would represent the tidal distortion induced by the MW to the outer North-East of the LMC disk.\par
To be complete it is worthwhile to underline that some models of the LMC interacting with the SMC and the MW do not expect the presence of a diffuse bended outer structure.
For example, \citet{Besla-2016} brought to light the existence of stellar arcs and multiple stellar arms to the North of the LMC. From hydrodynamical simulations of the MCs either when evolved in isolation or under the gravitational influence of the MW, they concluded that repeated close interactions with the SMC are the main cause of the many stellar substructures observed in the Northern periphery, included the NTA.
However, their models were based on the observations that a Southern counterpart of the features revealed to the North does not exist. These Southern substructures (S-SUB in Fig.~\ref{fig:gmm}) were disclosed afterwards by \citet{mackey2018} and \citet{Belokurov&Erkal2019}. Therefore, because of the asymmetric features revealed only to the North, \citet{Besla-2016} disfavor a formation scenario driven by global Milky Way tides. This is at odds with the more recent models carried out by \citet{Belokurov&Erkal2019} and \citet{Cullinane-2022} which are based on more recent and complete data on the features unveiled around the LMC.\par

\begin{figure*}
    \centering
    \includegraphics[width=0.48\textwidth]{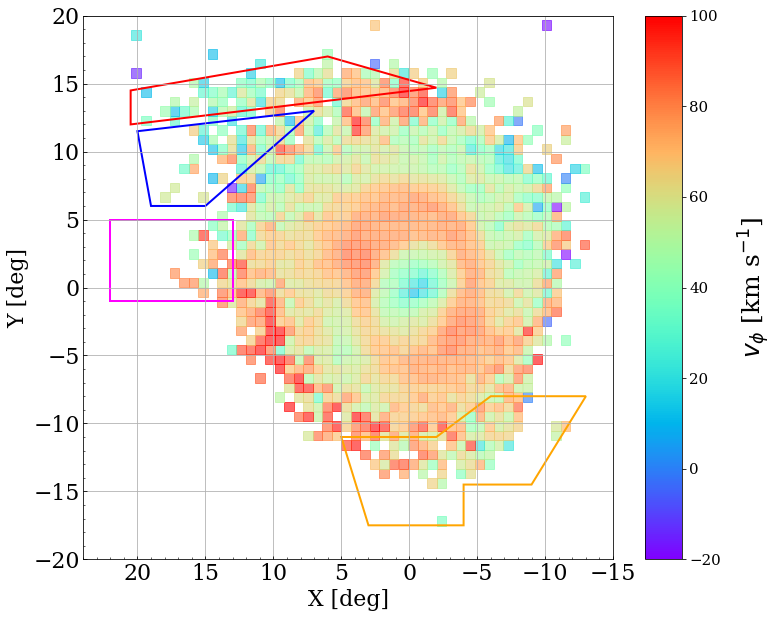}
    \includegraphics[width=0.48\textwidth]{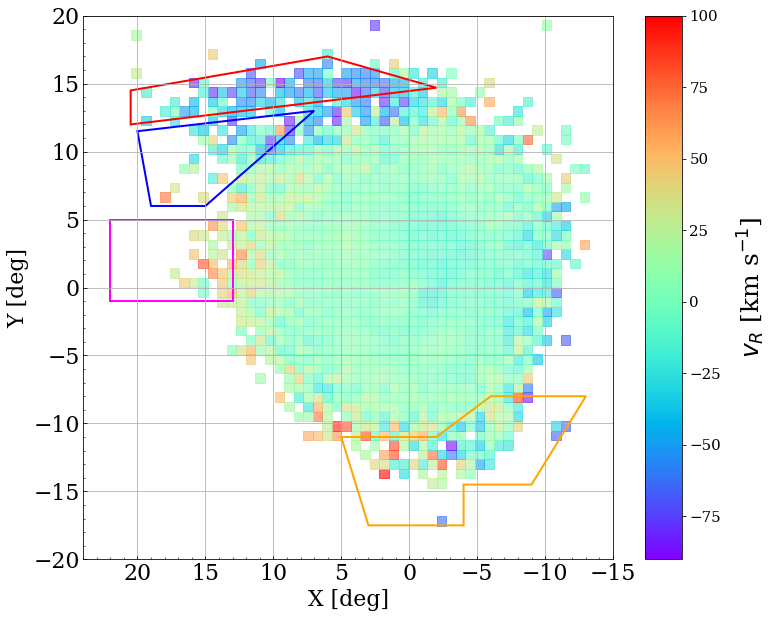}\\
    \includegraphics[width=0.48\textwidth]{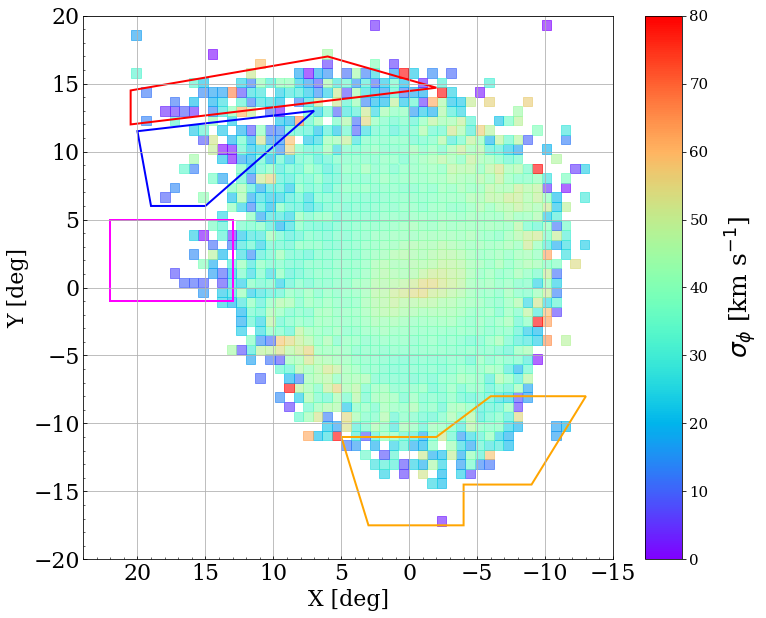}
    \includegraphics[width=0.48\textwidth]{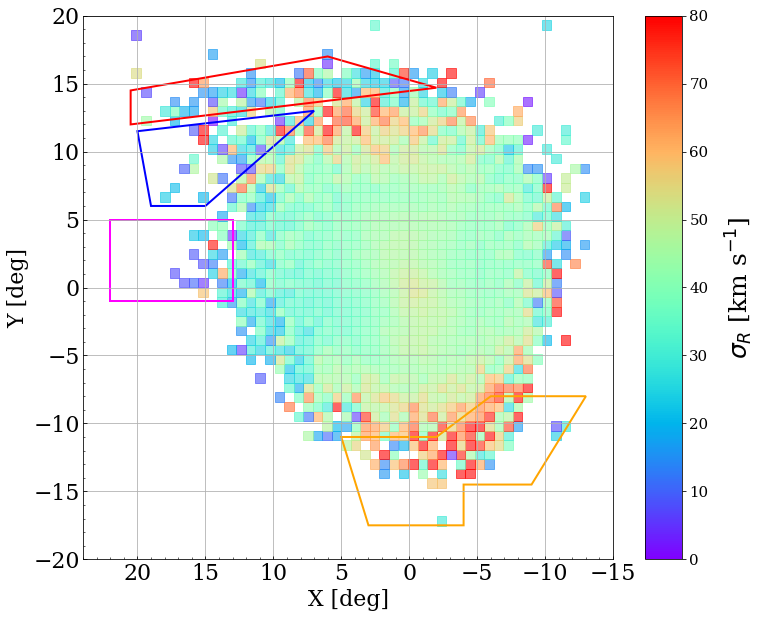}\\
    \caption{\emph{Top:} Median velocity of the LMC tangential velocity $v_{\phi}$ (left panel) and  LMC radial velocity $v_R$ (right panel). 
    \emph{Bottom:} Velocity dispersion of $v_{\phi}$ (left panel) and $v_R$ (right panel). 
    Each pixel is $0.7\degr \times 0.7\degr$ in size. The coloured polygons are the same as in the right panel of Fig.~\ref{fig:gmm}}
    \label{fig:lmc_PM}
\end{figure*}

\subsection{Velocity field}

If the NES and NTA had the same origin, they should also show a coherent velocity field.
In the context of the {\it MagEs} survey, \citet{Cullinane-2022} analyzed the kinematics of the NTA through spectroscopic measurements of different fields centred along the arm, in conjunction with {\it Gaia} EDR3 PMs. From a combined study of NTA metallicity and overall kinematics, they definitely confirmed that the NTA is made of LMC outer disc material tidally distorted by an external gravitational potential.
They also carried out a suite of N-body simulations, obtaining that the LMC embedded in a heavy-MW potential (mass MW $\sim 10^{12}$) represents the best qualitative match between models and observations.
However, their best model shows a diffuse stellar component connecting the LMC disc with the NTA, which resembles the NES.
From the dynamical point of view, the \citet{Cullinane-2022} simulations are not able to reproduce the measured kinematics of the NTA.
Indeed, while the measured tangential velocity is quite well replicated by the best model, the radial and vertical velocities are inconsistent with stars in an equilibrium disc and very different from what is expected by a solely LMC-MW interaction.\par 
To further investigate this point, Fig.~\ref{fig:lmc_PM} (top row) shows the LMC $v_{\phi}$ (left panel) and $v_R$ (right panel), calculated as median of the velocity of the stars within pixels of $0.7\degr \times 0.7\degr$, and their dispersions (bottom row). 
Both these diagrams seem to suggest a similar velocity pattern for the NTA and the NES, constituted by strong velocity gradients, with their velocity decreasing either going from the ES towards the NTA or moving along the NTA towards the West direction.
According to the velocity dispersion maps in the lower panels of Fig.~\ref{fig:lmc_PM} (the one in $\sigma_R$, in particular), NES and NTA appear dynamically colder than the adjacent edge of the LMC disc (around $(X,Y) \simeq (5\degr,12\degr)$ or than other tidal structures, such as S-SUB, which is very hot (i.e. $\sigma_R \geq 70~{\rm km~s^{-1}}$).
However, some of the pixels beyond $10\degr-15\degr$ include few stars, causing the outer velocity map of the LMC to be noisier and more scattered.
It is also visible from the $v_R$ plot that the NTA and the Finger present negative radial velocities (top right panel) of the order of $\simeq -50~{\rm km~s^{-1}}$, similar to those estimated by \citet{Cullinane-2022}, with peaks of $\simeq -80~{\rm km~s^{-1}}$.\par
To explain the origin of the peculiar observed radial velocities of NTA's stars \citet{Cullinane-2022} speculated that they could have been caused by recurring close passages of the SMC, happened more than 1 Gyr ago, beyond the time span of their models, which cover only the last Gyr. According to their simulations, any close passage of the SMC in this period was not able to alter significantly the kinematics of the NTA.
In this context, the Finger might have been affected by past nearby passages of the SMC as well, hence it might be connected to the NTA and physically distinct from the rest of the NES.
\begin{figure}
    \includegraphics[width=\hsize]{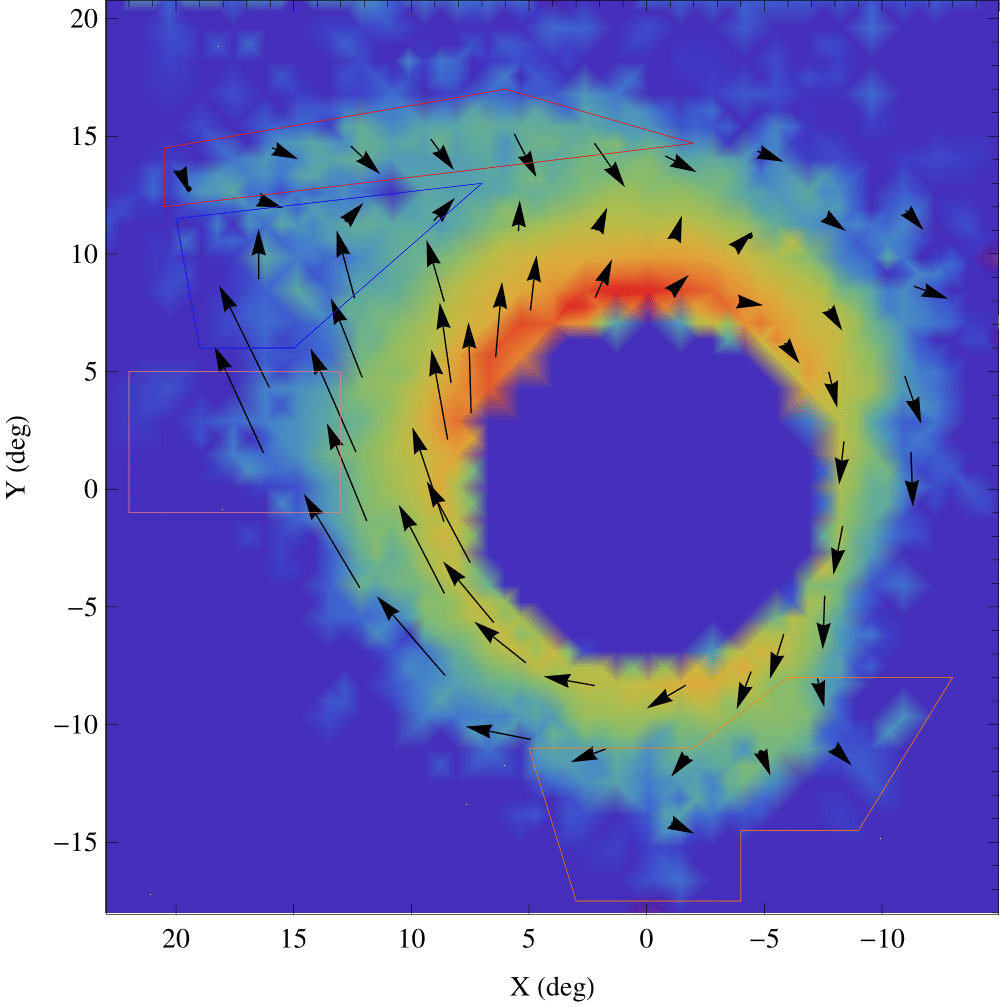}
    \caption{Velocity map (arrows) of the LMC overlaid on a star density map of the regions beyond 9\degr~from the LMC centre. Bin size is 0.7\degr $\times$ 0.7\degr. The length of the arrows is proportional to the velocity vector's module.}
    \label{fig:vel_vectors}
\end{figure}
However, the velocity maps of the NTA and NES displayed in Fig.~\ref{fig:lmc_PM} do not allow us to establish whether the NES-NTA is a single highly disturbed sub-structure or is made by two uncorrelated physical sub-structures originated at different times.
An additional characteristic visible in Fig.~\ref{fig:lmc_PM} is the large positive radial velocity shown by ES stars, with peaks of $v_R \simeq 80~{\rm km~s^{-1}}$, possibly indicating that the ES is in an ongoing stripping process.\par
An overall picture of the bulk motion of the outer regions, corrected for the solar reflex motion, is depicted in Fig.~\ref{fig:vel_vectors}. 
At first glance the plot suggests that the NTA is being pulled back by the gravitational potential of the LMC towards equilibrium, as already visible in Fig.~\ref{fig:lmc_PM}. The overall kinematics of the ES, and more generally of the outer East-North-East side of the LMC, indicates that this side is strongly moving away from the LMC. The NES, instead, shows a more chaotic kinematics, as some regions have a motion more similar to the ES while others have an opposite velocity direction. In the same figure the highly disturbed kinematics in the S-SUB is also visible.
The high perturbations induced by the MW is testified by the strong deviations from the rotational motion of the East LMC side with respect to its West side. 
However, we caution the reader that the LMC inclination may affect the velocities discussed above and the consequent results. A quantitative discussion of this scenario could be provided only when the velocity field is drawn also with spectroscopically inferred radial velocities.
Recently, also \citet{Choi-2022} revealed large PM residuals in the Southern disc with a sample of RC and RGB stars collected from Gaia EDR3, even though their analysis was limited to the inner 6\degr~from the LMC centre. They compared the observed LMC PMs with the outcome of numerical simulations of a LMC in interaction with the MW and SMC, putting constraints on the lookback time the last close encounter between the MCs happened (i.e., $t < 250$~Myrs ago) and on its impact parameter (i.e., $\leq 10$~kpc.)\footnote{These constraints are very similar to those inferred by \citet{Zivick2018}}.
\par
An additional interesting feature in the North-East side of the LMC was recently discovered by \citet{Petersen-2021} using the RR Lyrae variables as kinematics tracers. Indeed, they detected a tidally stripped stream, expected to be the outcome of the MW tidal force, in the same direction of the features we have discussed in the North-East side of the LMC. \citet{Petersen-2021} probed only regions beyond $20-25\degr$~from the LMC centre, and revealed the stream signal up to $\sim 30\degr$~(see their fig. 1), finding also likely stream members in the Northern hemisphere at about 70\degr~from the LMC centre.
In this context, we speculate that the ensemble of stellar streams in the North-East outer disc could represent the inner part of the outer stream disclosed by \citet{Petersen-2021}, which extends up to the Galactic disc.\par
Dynamical simulations of LMC-SMC-MW interactions, going back to at least 2-3 Gyr, are essential to explain the peculiar kinematic features observed in the North-East LMC, such as the strong velocity gradients along the NES-NTA sub-structures.

\section{Summary}

In this work we exploited the exquisite multi-dimensional data provided by {\it Gaia} EDR3 to investigate the outer low surface brightness substructures of the LMC.
We selected LMC RGB and RC stars based on their position on the CMD and filtered out all the stars having angular momentum beyond $5~\sigma$ that of the average LMC disc.
We run the unsupervised clustering technique GMM with 6 input parameters to separate the remaining stars into the three main stellar components, the MW, LMC and SMC, and we limited our analysis to highly-likely LMC members ($P >99\%$).\par
The GMM was able to recover the outer sub-structures discovered in the last years in the LMC, such as the NTA, the S-SUB or the ES. In addition,
this procedure allowed us to detect an unknown diffuse tidal structure at its North-East side extending up to $\sim 20\degr$ from the LMC centre, the North-East Structure (NES).
The NES fills the gap between the outer LMC disk and the other known structures in the North-East of the LMC, namely the ES and the NTA. 
The presence of tidally distorted LMC stars at the location of the NES was expected based on N-body simulations of a MW-LMC interaction \citep[][]{Belokurov&Erkal2019,Cullinane-2022} as a consequence of the tidal stress induced by the MW on the LMC. This occurrence might induce us to think that the NES has the same origin as the NTA. 
From the kinematic analysis based on the Gaia EDR3 PMs, these tidal stellar components reveal a strong velocity gradient towards the outer LMC, but the present dataset is not sufficient to assess whether or not the NES is physically correlated to the NTA. Overall, the North-East side of the LMC displays radial velocities not consistent with a disc in equilibrium, showing large negative velocities for the NTA-Finger and positive for the ES. This occurrence indicates that the outer LMC disc has been strongly disturbed in the past few Gyrs.
Spectroscopic follow-up to measure accurate radial velocities and more detailed and extended in time N-body simulations are requested to assess the origin of the NES and of the other sub-structures in the North-East side of the LMC.

\begin{acknowledgments}

We warmly thank the anonymous Referee for the comments that helped us to improve the manuscript. 
This work has been partially supported by INAF through the “Main Stream SSH program" (1.05.01.86.28).
M.G. and V.R. acknowledge support from the INAF fund "Funzionamento VST" (1.05.03.02.04).\\
This work presents results from the European Space Agency (ESA) space mission Gaia. Gaia data are being processed by the Gaia Data Processing and Analysis Consortium (DPAC). Funding for the DPAC is provided by national institutions, in particular the institutions participating in the Gaia MultiLateral Agreement (MLA). The Gaia mission website is https://www.cosmos.esa.int/gaia. The Gaia archive website is https://archives.esac.esa.int/gaia.

\end{acknowledgments}





%

\vspace{5mm}
\facilities{{\it Gaia} eDR3 \citep[][]{GaiaBrown2021}}


\software{{\rm TOPCAT} \citep{topcat}, scikit-learn \citep{scikit-learn}, matplotlib \citep[][]{matplotlib}, gala \citep[][]{gala_software}          }



\appendix

\section{Prior selection through Gaia}
\label{app:query}

In the following the query we made on the {\it Gaia} EDR3 dataset as described in Sect.~\ref{sec:sample}:

\begin{verbatim}
SELECT 
source_id,ra,dec,phot_g_mean_mag,phot_bp_mean_mag,phot_rp_mean_mag,bp_rp,parallax,parallax_error,
pmra,pmra_error,pmdec,pmdec_error,l,b,1.085736/phot_g_mean_flux_over_error,
1.085736/phot_bp_mean_flux_over_error,1.085736/phot_rp_mean_flux_over_error
FROM gaiaedr3.gaia_source    
WHERE
abs(parallax)<3.0*parallax_error
AND
l>200 AND l<360 AND b<-10 AND b>-90
AND
pmra>-0.5 AND pmra < 3.0 AND pmdec > -2.2 AND pmdec < 2.2 
AND
pmra_error<0.2 AND pmdec_error<0.2
AND
phot_bp_rp_excess_factor < 1.3+0.06*power(phot_bp_mean_mag-phot_rp_mean_mag,2)  
AND
phot_bp_rp_excess_factor > 1.0+0.015*power(phot_bp_mean_mag-phot_rp_mean_mag,2) 
AND
ruwe<1.3
AND
astrometric_excess_noise_sig<2


\end{verbatim}

The following describes the polygon edges used to select RGB and RC stars by means of {\rm TOPCAT}:

\begin{verbatim}
isInside(bp0-rp0, g0, 1.0, 18.35, 1.1, 17.5, 1.25, 16.7, 1.65, 15.50, 2.1,
15.50, 1.6, 16.60, 1.35, 17.65, 1.15, 18.9, 0.9, 18.90, 0.9, 18.35)
    \end{verbatim}

\section{Density maps of the outer LMC}
\label{app:kde}

In this section we show the density map of the LMC as depicted in Fig.\ref{fig:gmm}, with different choices of bin size and width of the kernel function. To enhance the small substructures we adopted narrower kernel function, while to have a global view of the outer LMC we set a larger bin size and shallower kernel function.

\begin{figure}
    \centering
    \includegraphics[width = 0.49\hsize]{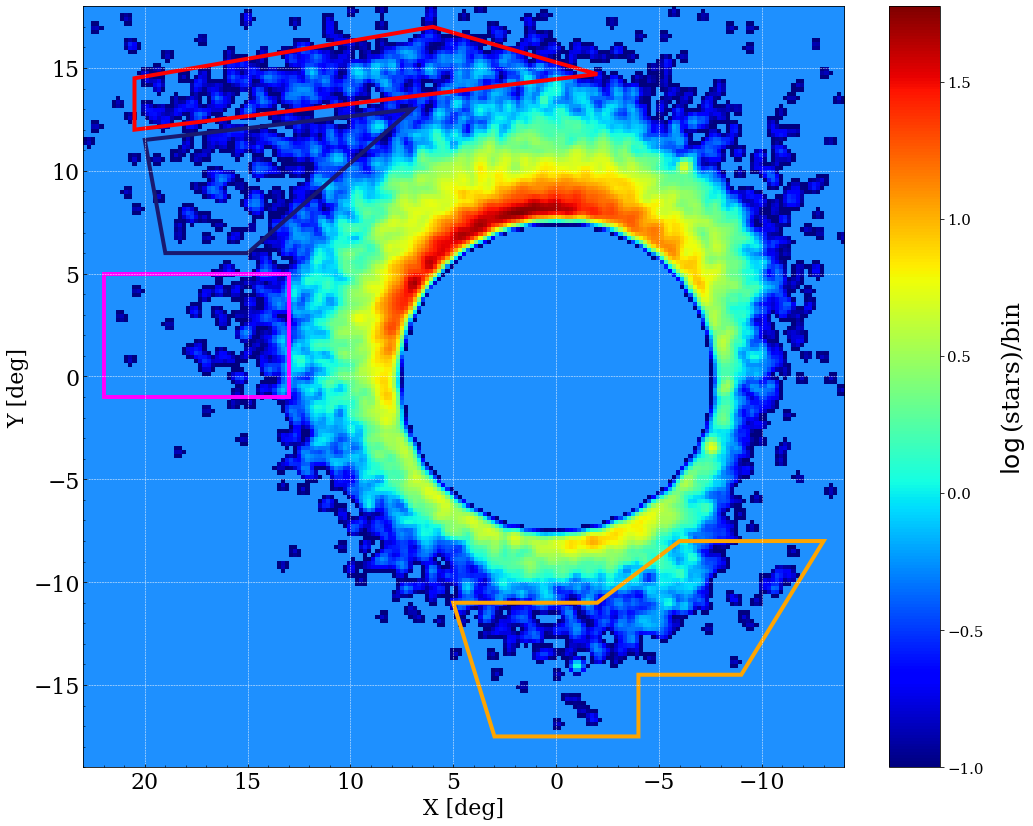}
    \includegraphics[width = 0.49\hsize]{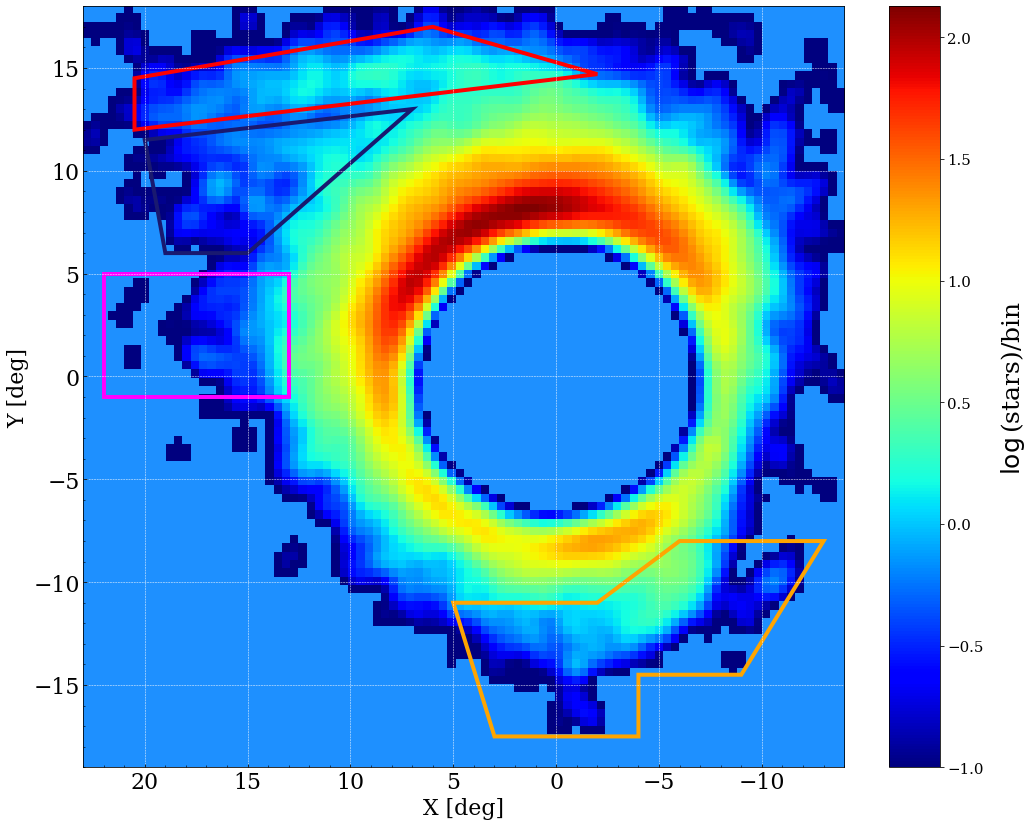}\\
    \caption{Same as in Fig.\ref{fig:gmm}, but with a bin size of $0.2\degr \times 0.2\degr$ smoothed with a Gaussian kernel of $\sigma = 0.2\degr$ (top panel) and a bin size of $0.4\degr \times 0.4\degr$ smoothed with a Gaussian kernel of $\sigma = 0.5\degr$ (bottom panel).}
    \label{fig:kde_LMC}
\end{figure}



\bibliography{mybibliography_sbp}{}
\bibliographystyle{aasjournal}






\end{document}